# A systematic review on visual-processing deficits in Neurofibromatosis type 1: what possible impact on learning to read?

Marie Vernet[a,b,c], Stéphanie Ducrot[a], Yves Chaix[c,d]

[a] Aix Marseille Univ, CNRS, LPL, Aix-en-Provence, France

[b] Centre de jour enfants, Centre hospitalier de Digne-les-Bains, France

[c] ToNIC, Toulouse NeuroImaging Center, Université de Toulouse, Inserm, UPS, Toulouse, France

[d] Hôpital des enfants, Centre hospitalier universitaire Purpan, Toulouse, France

SHORT RUNNING TITLE: VISUAL-PROCESSING SKILLS IN NF1 CHILDREN

## Correspondance

Marie Vernet
Centre de Jour, Site Romieu
Rue de l'ancienne maternité
04000 Digne-les-Bains, France
Email: mvernet@ch-digne.fr

### ACKNOWLEDGEMENTS

The authors are very grateful to the "Association Neurofibromatoses et Recklinghausen" (France) and the "Fondation de France" for their support and financial contributions to this project. We are also grateful to N. Sghaier for her valuable information and advice.

### FUNDING

This work was funded by a PhD Fellowship from the "Fondation de France" and the "Association Neurofibromatoses et Recklinghausen" provided to the first author (VISALECT_NF1_00099576).

### DISCLOSURE STATEMENT

The authors have no conflicts of interest to disclose.

**ABSTRACT**

This systematic review aimed to examine the possible implication of visual-perceptual, visuo-attentional and oculomotor processing in the reading deficits frequently experienced by children with Neurofibromatosis type 1 (NF1), as previously shown in dyslexia. Using PRISMA methodological guidelines, we examined 49 studies; most of these reported visual-processing deficits in this population, raising the importance of directly studying the visuo-perceptual and visuo-attentional processes and eye-movement control involved in the learning-to-read process in NF1. The discussion provides a reflection for a better understanding of how visual-processing skills interact with reading deficits in NF1, as well as new avenues for their screening and care.

## INTRODUCTION

Neurofibromatosis type 1 (NF1) is a neurogenetic disorder, affecting approximately 1 in 2 500 to 3 000 births (Evans et al., 2010). The diagnosis is based on the clinical criteria defined in the recently revised version of the NIH Consensus Conference Statement (Legius et al., 2021; National Institutes of Health, 1988). The associated medical manifestations are multiple, affecting cutaneous, ophthalmologic, orthopaedic as well as neurologic and cognitive domains (Baudou & Chaix, 2020). A prominent feature of this disease concerns the heterogeneity of the affected individuals' profiles, either on the medical or cognitive levels. In that sense, cognitive deficits can affect attention, language, executive functions, praxis or even visual-perceptual processes (Hyman et al., 2005; see also Crow et al., 2022; Lehtonen et al., 2013, for reviews). All these impairments could explain the high frequency of learning disabilities found in this population. Specifically, 30 to 60% of children with NF1 exhibit difficulties in reading acquisition (Descheemaeker et al., 2005; Hyman et al., 2006; Orraca-Castillo et al., 2014). Although this prevalence varies greatly between studies, it is always higher than that observed in the general population, where nearly 10% of children and adolescents experience reading difficulties (Andreu et al., 2021; Chabanon, 2021). Understanding the mechanisms underlying reading difficulties in NF1 has therefore been the focus of ongoing research, with particular interest in the role of visuo-perceptual and visuo-attentional processes, and eye-movement control in the learning-to-read process.

In NF1 children, reading deficits include deficits in single-word and non-word reading efficiency (Arnold et al., 2020; Cutting et al., 2000; Hyman et al., 2005; Lehtonen et al., 2015; Orraca-Castillo et al., 2014; Watt et al., 2008). At the text level, difficulties can affect reading speed, accuracy (Chaix et al., 2017) and also comprehension (Arnold et al., 2020; Biotteau et al., 2019; Cutting et al., 2000; Orraca-Castillo et al., 2014). In this context, the graphophonological decoding of written words appears to be greatly impacted: two studies of native English-speaking children with NF1 have shown that between 49% and 67% of these children presented word decoding problems (Arnold et al., 2020; Watt et al., 2008). These may be explained in part by the phonological impairments found in this population and considered as an integral feature of the neurocognitive profile of this disease (Arnold et al., 2018; Chaix et al., 2017; Cutting et al., 2000; Cutting & Levine, 2010). More generally, linguistic skills known to be related to reading, such as phonological awareness, phonological memory, rapid automatic naming (RAN), and letter-sound knowledge were also found to be impaired in children with NF1 (Arnold et al., 2018; Cutting & Levine, 2010). Notably, the deficit in the RAN process observed in NF1 children with a reading deficit was not corroborated in the whole NF1 group when the reading level was not taken into account (Cutting et al., 2000; Mazzocco et al., 1995). The deficit in RAN in NF1 seems to be specifically associated with reading difficulties.

Although phonological processing is a necessary (but not sufficient) condition for the development of adequate word recognition skill, our ability to read depends also on visuo-attentional processes and eye movement control (e.g., Bellocchi et al., 2017; Ducrot et al., 2013; Facoetti et al., 2010; for reviews, see Hung, 2021; Premeti et al., 2022). Reading does require children to focus selectively on words using left-to-right attentional scanning. The high visual acuity needed to rapidly identify words is spatially limited and beginning readers have to learn to move their eyes in order to optimize the processing of the majority of words in the text being read (Grainger, 2018; O'Regan & Lévy-Schoen, 1987; Rayner, 1986). The effectiveness of oculomotor control emerges when learning to read, with an improvement in eye movement parameters. For instance, duration, number of fixations and number of regressions decrease, while the amplitude of saccades and the probability of skipping a word increase (e.g., Blythe et al., 2009; Ducrot et al., 2013; Lopukhina et al., 2022; Vorstius et al., 2014). As children learn to read and take the left-to-right directionality of reading into account, a left-right visual field (VF) asymmetry emerges. The implementation of an optimal saccade targeting strategy early in the learning-to-read process explains in part the enhanced efficiency of these eye movement parameters, with average initial landing positions gradually shifted towards the left of the word's center (preferred viewing location, PVL) during the first years of formal reading instruction (Ducrot et al., 2013; Huestegge et al., 2009; Rayner, 1979). Improved efficiency of eye movement control is related to the increase in the size of the perceptual span[1] during development (extending asymmetrically in the direction of reading, even in beginning readers; Häikiö et al., 2009). In this context, the distribution of visual-spatial attention in foveal and parafoveal areas plays an essential role. The voluntary allocation of attentional resources in the fovea supports the processing and accurate identification of written words. It was defined by a visual attention span which corresponds to the number of orthographic units that can be processed simultaneously in the foveal regions in one fixation (Bosse et al., 2007; Bosse & Valdois, 2009)Erreur ! Signet non défini.. This visual attention span is related to reading skills regardless of phonological awareness processes (Liu et al., 2022; Lobier et al., 2013; Valdois et al., 2019). In addition to the foveal allocation of attention, pre-attentive processing in the parafovea is also important as it facilitates the processing of the currently fixed word but also influences saccadic computation towards the next fixation (for a review, see Schotter et al., 2012).

[1] For a discussion on the differences between perceptual, visual and visual-attentional spans see Frey and Bosse (2018).

Furthermore, visual processing shapes the way visual information is extracted from print (Aghababian & Nazir, 2000; Ducrot et al., 2013). The deployment of attentional resources on visual information modulates the level of perceptual processing[2] implemented. For instance, expert readers spread their attentional focus over the whole word to be processed, and demonstrate a global precedence on local perception (Austen & Enns, 2000; Krakowski et al., 2015, 2016, 2018; Navon, 1977; Poirel et al., 2014). During development, qualitative changes in perceptual analysis occur with a local precedence in preschoolers and then a global precedence implemented from 6-7 years (Krakowski et al., 2016, 2018; Poirel et al., 2008; Porporino et al., 2004; Schmitt et al., 2019). This change in the level of perceptual analysis from the age of 6-7 years onwards corresponds to the beginning of explicit reading instruction in the first grade (Krakowski et al., 2016; Poirel et al., 2008), with a switch from a grapho-phonological decoding mode (letter-by-letter processing), to an orthographic processing mode (characterized by an optimal viewing position for word processing). This is an argument for the already demonstrated involvement of global-to-local levels of perceptual analysis in reading (Franceschini et al., 2017, 2021). This hypothesis is consistent with the idea that the level of perceptual analysis depends on the spatial distribution of attentional resources, according to the increase in the visual attention span in parallel with the reading level (Bosse & Valdois, 2009). With expertise, the attention deployed on words is adapted to the preferred global processing mode (Ans et al., 1998) and is related to the efficiency of the oculomotor parameters (Prado et al., 2007).

[2] The level of perceptual processing was introduced by Navon (1977), using cross-level hierarchical stimuli to study (1) the primacy of one level over the others and (2) the factors affecting this process.

Orthographic identification is also influenced by multimodal learning in which visual-motor processes are involved. Several studies have shown that visual-motor processes assessed through tasks involving grapho-motor response are related to reading abilities (Bellocchi et al., 2017; Hopkins et al., 2019; Meng et al., 2019). This multimodal learning allows the integration of both the visual configuration of the stimulus and its motor execution pattern, explaining why visual-motor skills are important for learning to read (Suggate et al., 2018). In this sense, Longcamp et al. (2005) reported that preschoolers' training in handwriting induced better letter recognition than typing, strengthening the idea of visual-motor involvement in reading.

It follows that basic aspects of oculomotor control (which provide optimal visual input), the ability to orient the focus of attention as well as the ability to control its size are assumed to play a crucial role in the development of reading skills (Ducrot et al., 2013; Grainger, 2018; Leibnitz et al., 2017; Morris & Rayner, 1991). Accordingly, they were shown to longitudinally predict future poor reading skills and to differentiate children with developmental dyslexia (DD) from typically developing children (TD; e.g., Bellocchi et al., 2017; Franceschini et al., 2012, 2017; Giovagnoli et al., 2016; Meng et al., 2019; Son & Meisels, 2006; Vernet et al., 2022a). Note however that there is considerable debate in the literature about whether these visual deficits play a causal role in dyslexia or whether they reflect an underlying deficit in the processing of written words (Blythe et al., 2018). Poor readers and individuals diagnosed as having dyslexia exhibit inefficient eye-movement patterns with longer fixation times, shorter saccade amplitudes, more regressions and fewer skipped words in eye-tracking studies compared to expert readers (De Luca et al., 2002; Franzen et al., 2021; Hawelka et al., 2010; Lefton et al., 1979). However, the attempts to replicate differences in eye-movement patterns of dyslexic, poor and control readers when performing non-reading tasks have been essentially unsuccessful, suggesting that eye movements are functional and perform the same function in DD readers as they do in proficient readers. The point of agreement is that in a reading task, children with DD exhibited atypical visual-processing skills. Among them, the deployment of visual attention seems to be impaired in DD individuals. It has been reported that individuals with DD demonstrate a more distributed/diffused mode of attention and have difficulty in narrowing their focus of attention (Facoetti et al., 2000, 2003; Franceschini et al., 2012), resulting in occasional oculomotor control deficits and a diffuse spread of initial landing positions. In addition, in parafoveal vision, an abnormally important allocation of attentional resources in the right VF was found in dyslexia and is associated with a left VF "mini-neglect" and a right VF over-distractibility (e.g., Facoetti & Turatto, 2000; Geiger et al., 2008; Lorusso et al., 2004). This filtering defect accounts for the great sensitivity to visual crowding[3] experienced by DD children (for a review, see Bellocchi, 2013). Finally, the atypical eye-movement patterns observed in dyslexic children have been linked to impaired visuo-attentional processing in foveal vision, and poor visual attentional span abilities, resulting in fewer letters simultaneously processed and more rightwards fixations (Prado et al., 2007). This idea is consistent with the atypical perceptual analysis observed in children with dyslexia, depending excessively on the local level of processing with a defect in global processing compared to typical readers (Bedoin, 2017; Franceschini et al., 2017; Schmitt et al., 2019). As suggested by Facoetti (2012), a possible neurobiological substrate of visuo-spatial attention deficits in DD could be weakened or abnormal magnocellular input to the dorsal visual stream. Magnocellular–dorsal deficits could lead to reading difficulties through impaired serial attentional orienting (Facoetti et al., 2010; Franceschini et al., 2012; Vidyasagar & Pammer, 2010) or poor eye-movement control (Stein, 2001, 2019, but see Goswami, 2015, for a different point of view).

Summing up, a high proportion of children with NF1 experience a reading deficit. In this context, the implication of linguistic skills in reading difficulties has previously been highlighted in NF1 children. Visuo-attentional processes specific to reading behaviour are assumed to be highly essential in the development of reading skills and that their inadequate development might be one cause of reading disabilities. However, despite the frequency of reading deficits in NF1 children, very few studies have directly investigated the involvement of visual-processing difficulties in reading learning failure in NF1. However, reading difficulties constitute a critical area of interest given their lifelong negative implications (see Livingston et al., 2018, for a review on DD). Moreover, nonverbal learning disabilities and more precisely visual-spatial perception deficits have long been defined as a hallmark feature of the NF1 neurocognitive profile (Eldridge et al., 1989; Eliason, 1986). In light of all these observations, this systematic review aimed to examine the literature on visuo-perceptual processing (with and without motor implications) and visuo-attentional processes in children with NF1, and to investigate the possible impact of these processes on the frequent reading deficits observed in these children. To this end, the present review targets data from neuropsychological assessments, experimental tasks, neuroimaging, and eye-tracking to study the efficiency of these visual processes.

[3] The visual crowding effect corresponds to the identification of a visual target being harder due to interference from surrounding visual objects, compared to when it is isolated.

## METHOD

This systematic review was based on the PRISMA (Preferred Reporting Items for Systematic Reviews and Meta-Analyses) methodological guidelines (Gates & March, 2016; Page et al., 2021).

***Eligibility criteria.*** Studies included in the present systematic review were required to meet all of the following eligibility criteria: (1) articles were original studies published before January 2023, (2) patients included met the clinical diagnostic criteria for NF1, according to the Neurofibromatosis Conference Statement (National Institutes of Health, 1988), (3) participants were children and adolescents with NF1 younger than 18 years old, (4) studies assessed visual processes, specifically visuo-perceptual, visuo-attentional or oculomotor processes, (5) participants did not experience any other neurological disease in addition to NF1, and finally, (6) studies were English language full text.

***Information sources and search strategy.*** The literature search was performed in March 2023[4], from three computerised databases: PubMed, Embase and PsycARTICLES. The search for articles was carried out using the following combination of keywords: (“NF1” or “neurofibromatosis type 1”) and (“vision” or “attention” or “visuo-perceptual” or “visuo-attentional” or “oculomotor”) and (“cognitive” or “neuropsychology” or “neurocognitive”). No other filters were applied at this stage of the paper search to avoid missing studies relevant to the current research question. For the same reason, additional studies identified through other sources (e.g., citation searching) were included in the selection process.

[4] Note that the literature search on the Embase database was conducted in March 2022 considering papers before January 2022. This search could not be updated in 2023 due to access rights to this database.

***Selection process.*** A single reviewer (M. V.) carried out the article selection process to ensure consistency between the different screening steps. The first assessment phase was to eliminate duplicate articles. Then, the studies were checked for eligibility by examining their titles and abstracts, and finally, by a careful reading of the remaining articles' full texts. All eligibility criteria mentioned above were verified during the selection process. At the same time, we decided to exclude articles that specifically focused on the effect of drugs or disorders associated with NF1 (e.g., Attention Deficit Hyperactivity Disorder, ADHD or Autism Spectrum Disorder, ASD) on cognitive functioning, and that did not directly evaluate the effect of NF1 on cognitive performance. Given the large proportion of ADHD in this population (nearly 40% of children with NF1, Hyman et al., 2005), the inclusion of children with NF1+ADHD in a study was obviously not an exclusion criterion for the current review. However, studies focusing specifically on the involvement of ADHD in NF1, based on a specific methodology that did not allow us to report and account for the effectiveness of visual processes in NF1 were excluded. We also removed papers focusing only on sensory impairments of vision and studies reporting results on the same cohort as another article and not providing additional information on the present issue. For the psychometric tests, we retained only the studies involving tests that assessed visual processing in 2D, whether using a paper-pencil or computerised version, to be as close as possible to the conditions found in reading. The tests eligible for inclusion in the reviewed studies and the visual processes they assess are listed in Table 1.

-------------------------------------------------------------------

[Insert Table 1 about here]

[Insert Figure 1 about here]

-----------------------------------------------------------------

## RESULTS

The search strategy resulted in 516 articles, of which 49 met the criteria for inclusion in this systematic review. The flow diagram of the study selection process is detailed in Figure 1 and the studies' characteristics are reported in Table 2. The results are organized according to the visual processes studied: visuo-perceptual processes with a reduced motor implication (i.e., visuo-spatial perception, global-to-local visual processing and visual pathways functioning), perceptual-motor processes (i.e., visual-motor abilities and saccadic system functioning) and finally, visuo-attentional skills. The same study can be included in different sections if they assessed several of our interest processes. NF1 performance compared to the control groups for the different visual processes of interest in the selected studies are schematically represented in Table 3.

-----------------------------------------------------------------

[Insert Table 2 about here]

[Insert Table 3 about here]

-----------------------------------------------------------------

### *Visuo-perceptual processing*[5].

*Visuo-spatial perception.*

Visuo-spatial perception in NF1 children was extensively studied, in particular using the *Judgment of Line Orientation test (JLO)*. The majority of these studies showed poorer performance on this task in NF1 children compared to siblings, TD subjects and also normative data (Arnold et al., 2020; Barquero et al., 2015; Baudou et al., 2020; Chaix et al., 2017; Clements-Stephens et al., 2008;

[5] Note that this section includes only visuo-perceptual processes with reduced motor involvement. Perceptual-motor processes will be discussed in a subsequent section.

Cutting & Levine, 2010; Dilts et al., 1996; Eldridge et al., 1989; Erdogan-Bakar et al., 2009; Gilboa et al., 2014; Hofman et al., 1994; Hyman et al., 2005; Isenberg et al., 2013; Krab et al., 2008; Lehtonen et al., 2015; Mazzocco et al., 1995; Moore et al., 2000; Payne et al., 2013; Ribeiro et al., 2012; Ullrich et al., 2010; Watt et al., 2008; but see Billingsley et al., 2002; Cutting et al., 2000, for non-significant difference). Moreover, the proportion of children performing below the mean on the JLO test is significantly higher in NF1 compared to TD children (Dilts et al., 1996; Eldridge et al., 1989). Hyman et al. (2005) showed for instance that 56.3% of NF1 children scored more than one standard deviation below the mean on the JLO, compared to only 14.6% of controls.

Significantly worse visuo-spatial outcomes than controls were also shown on other tests such as: the *Thurstone test (*Chaix et al., 2017), the *TVPS* (Arnold et al., 2018; Dilts et al., 1996; but see Sangster et al., 2011 for a non-significant result on the *Visual discrimination* subtest) and also the *Spatial relations* subtest (*WJ-R*), the *Gap matching* and the *Line orientation* subtests (*BORB*; Hyman et al., 2005). However, no significant difference between groups was found on the *Recognition-discrimination* subtest (*FKSB*) in either the Billingsley et al. (2002) or Moore et al. (2000) studies. The same result was reported on the *HVOT* (Clements-Stephens et al., 2008; Cutting & Levine, 2010). Finally, visuo-spatial abilities assessed with the *Position in space* subtest of the *DTVP* did not lead to a consensus: while Clements-Stephens et al. (2008) showed poor scores for NF1 participants compared to control, no significant difference emerged in Mazzocco (2001) study.

An important issue concerns the IQ implication in the low scores of NF1 children. Payne et al. (2013) revealed a strong relationship between IQ and JLO outcomes. In line with this observation, Roy et al. (2010) reported that, after controlling for IQ, the lower scores of NF1 children compared to controls on the *Arrows* subtest *(NEPSY)* did not remain significant. Conversely, other studies showed significantly reduced visuo-spatial perception skills remained in the NF1 group even after taking into account IQ as a covariate (Hyman et al., 2005; Krab et al., 2008). Regarding all these results, one limitation needs to be stressed. In these studies, the index considered was the full-scale IQ, which includes a visuo-perceptual component. This could reduce, or even remove, the differences observed previously and must therefore be taken into account when interpreting these results.

Regarding the relationship between visuo-spatial perception abilities and reading proficiency, very few studies were conducted. The results on this issue showed lower visuo-spatial abilities obtained in NF1 children with a reading deficit compared to children with reading disabilities without NF1 (RD) and TD children (Barquero et al., 2015; Cutting & Levine, 2010; see D'Archangel et al., 2022, for results with trend significance between NF1 and non-NF1 children both with reading deficits). Cutting and Levine (2010) also demonstrated a significant association between word reading outcomes and visuo-spatial processes on the *JLO test* and the *Position in space* subtest. This link was demonstrated in NF1 children with a reading deficit, but not in TD, RD children, nor in NF1 children without a reading deficit. The last result could explain why Arnold et al. (2020) and Watt et al. (2008) did not find any significant correlation between the *JLO* performance and word reading in the whole NF1 group. In fact, not taking into account children's reading level could mask the significant association in NF1 children with reading difficulties because of the lack of association in NF1 children without reading difficulties (but see Mazzocco et al., 1995 for significant results). Another argument concerns preliteracy skills. Preschoolers with NF1 showed lower visuo-spatial abilities than controls and their performances were related to the spelling measure (Arnold et al., 2018). This supports the idea that visual deficits may contribute to the high frequency of literacy difficulties in NF1.

Finally, neuroimaging observations revealed a greater left than right hemisphere activation in NF1[6], whereas controls showed the opposite in a visuo-spatial perception task, even after controlling for IQ (Clements-Stephens et al., 2008). This supports the idea that in this neurodevelopmental disease there is an inefficient right hemisphere network related to deficient visuospatial processing. Note that additional activation was shown in frontal regions, which can be explained by the involvement of executive functions, either as an attempt to compensate for, or by their main role in visuo-spatial deficits (Roy et al., 2010).

*Global–Local visual processing.*

[6] Note that lateral dominance was not taken into account in this study, although structural changes may occur in NF1.

A disruption in the global-to-local visual processing seems to be a central feature of the visual processing difficulties reported in NF1 children. Payne et al. (2017) used a modified Navon paradigm[7] to study this issue (Navon, 1977). They replicated the expected global processing bias in control subjects, with no effect on congruency when they process the global form but a significant interference with incongruency when they process the local level of the letter. NF1 children, however, didn't show this global processing bias: they experienced significant interference when naming both local and global levels of stimuli. Thus, while the same high interference was highlighted between the two groups of children in local processing, the interference in global processing was significantly higher for children with NF1. In the same vein, Bulgheroni et al. (2019) suggested a predominance of local over global processing compared to controls in the *Rey-Osterrieth Complex Figure (ROCF)*. Although this test involves different processing skills, visual, executive, and motor functions, it is interesting to note that NF1 children favoured a more detail-oriented copy strategy rather than processing the whole shape of the figure, even though global processing is more relevant to the task.

The recognition of fragmented objects/shapes constitutes another way to evaluate global-to-local visual processing. The most commonly used in studies on NF1 is the *Visual closure* subtest (i.e., *DTVP*). While Mazzocco (2001) didn't find any difference between NF1 and TD children (as for *Figure-ground* and *Form constancy* subtests), the majority of the studies using this task showed poorer performance in the NF1 group (Barquero et al., 2015; Clements-Stephens et al., 2008; Cutting & Levine, 2010). Similarly, Van Eylen et al. (2017) pointed out that NF1 children need more contour information to recognise an object than TD children. However, this effect did not remain significant when executive functions were included as covariates. Note that the experimental task of Van Eylen et al. (2017) study, unlike the *Visual Closure* subtest, involves semantic knowledge of objects and requires access to them.

The implication of global-to-local visual processing in reading was not, to our knowledge, directly investigated in children with NF1. However, two studies demonstrated that NF1 children with a reading deficit performed poorly than TD children and non-NF1 children with a reading deficit on the *Visual closure* subtest (Barquero et al., 2015; Cutting & Levine, 2010). Note that NF1 children without a reading deficit performed halfway between the NF1 children with a reading deficit and the two control groups with no significant difference in performance with these groups. Finally, a significant correlation was also shown between single-word reading and *Visual closure* subtest outcomes but only in NF1 with a reading deficit (Cutting & Levine, 2010).

[7] In this task, the stimuli correspond to a large letter (global shape) composed of smaller letters (local shape). The congruency between the global and the local letters is manipulated. Participants are instructed to name the letter at a specific level of visual processing, either at the global or local level.

*Visual pathways involvement.*

The functioning of the visual processing pathways was first explored through the study of visual evoked potentials (VEPs). Iannaccone et al. (2002) showed that nearly 62.5 % of children with NF1 displayed abnormal VEP responses and specifically, that 44 % of the children presented an absence or a delay in the P2 component of the VEP in a visual flash stimulation, thus suggesting that NF1 children frequently demonstrated a primary abnormality of visual processing. Lalancette et al. (2022) also found reduced steady-state VEP responses in NF1 compared to controls. Finally, Ribeiro et al. (2014) demonstrated abnormal long-latency VEPs with a lower amplitude of the negative potential occurring at 300 ms than controls after chromatic visual stimulation. Note that this type of stimulation preferentially activates the parvocellular visual processing pathway.

Consistent with the findings from the VEP methodology, some studies suggest a delay in the maturation of the two main low-level visual pathways. Ribeiro et al. (2012) showed significant alterations of visual magnocellular and parvocellular systems with contrast sensitivity deficits for the achromatic higher spatial frequency, the achromatic low spatial high temporal frequency, and the chromatic red-green contrast sensitivity. Violante et al. (2012) corroborated these results in fMRI showing impaired activation of the low-level visual cortex in NF1 children, for the two kinds of stimuli activating respectively the magnocellular and parvocellular pathways. Contrary to these studies, Van Eylen et al. (2017) did not find a significant difference between individuals with NF1 and TD children in a coherent motion task that engages the activation of the magnocellular pathway. In the same vein, it is important to highlight that Ribeiro et al. (2012) failed to find correlations between the contrast sensitivity deficit found in NF1 and the neuropsychological measures, including notably the visuo-spatial perception abilities assessed by the JLO test.

***Perceptivo-motor processes*[8].**

*Visual-motor and visual-constructive skills.*

No consensus was reached on visual-motor processes regarding their preservation or impairment in NF1. Using the *Beery–Buktenica Developmental Test of Visual-Motor Integration (Beery VMI),* which is the one most commonly used in research on NF1, some studies demonstrated lower performance in NF1 children compared to controls (Dilts et al., 1996; Gilboa et al., 2010, 2014; Krab et al., 2008, 2011; Lorenzo et al., 2013)[9], while in some others no significant difference emerged between groups (Eldridge et al., 1989; Mazzocco, 2001; Moore et al., 2000; Sangster et al., 2011). Similarly, no agreement was reached with other tests. Using the *Bender visual-motor Gestalt* test, Erdogan-Bakar et al. (2009) did not find differences between individuals with NF1 and those with TD. Conversely, Casnar et al. (2014) showed significantly lower performance of NF1 children than the normative data and the control group on *Copying (DAS-II)* and *Visuo-motor precision (NEPSY-II)* subtests, with large (d = 1.33) and medium (d = 0.66) effect size respectively. The effect size for VMI corroborated the findings of Sangster et al. (2011) in preschoolers, since even after controlling for IQ and maternal educational level, no significant difference between groups remained. Several studies also assessed visual-constructive abilities through the *ROCF* test and all of them demonstrated poorer performance on figure copying in NF1 children compared to controls (Bulgheroni et al., 2019; Gilboa et al., 2014; Hofman et al., 1994; Mazzocco et al., 1995).

The previous results provided evidence that differences in the visual-motor and visual-constructive abilities of NF1 and TD children were not consensual across studies and tests. Even though there is no formal agreement, these processes have been frequently reported to be impaired in NF1. Indeed, between 23% and 65% of the NF1 children exhibited difficulties in the copying process (i.e., scores 1 SD below the mean; Casnar et al., 2014; Descheemaeker et al., 2005; Dilts et al., 1996; Hyman et al., 2005; Vaucheret Paz et al., 2019). It should be noted that despite the great variability of the prevalence between studies, it remained consistently high.

[8] In this section, a distinction was made between visual-motor and saccadic processes since these terms refer to two distinct literatures. The term visual-motor is specifically used in the context of grapho-motor tasks involving eye-hand coordination such as figure copying and line tracing between contours tasks.

[9] Note that the results provided by Gilboa et al. (2010) and Gilboa et al. (2014) used the same cohort and the same results in the two studies but comparing NF1 children with control subjects in one study and with normative data in the other. The same approach was performed in Krab et al. (2008) and Krab et al. (2011).

As with visuo-spatial perception processes, a point of divergence concerns the involvement of other cognitive processes in the tasks which could partly explain the failure to complete them. Indeed, Van Eylen et al. (2017) highlighted differences between NF1 and TD in the *ROCF* copy, which were no longer significant when executive functions with *flanker task* were included as covariates. On the other hand, Hyman et al. (2005) noted that spatial planning did not contribute significantly to the variance in the *ROCF* outcomes in NF1, whereas the *JLO* test predicted 26.3% of this variance.

Some researchers tried to link visual-motor skills with reading abilities but did not find significant correlations (Mazzocco, 2001; Watt et al., 2008). More generally, regarding school learning, Krab et al. (2008) identified differences between NF1 children with general learning disabilities and those with specific or no learning disabilities. The first group exhibited poorer *ROCF* and *VMI* scores than the two others. However, we could not define what was part of the reading deficit or other learning deficits since they were not dissociated in this study.

*Saccadic system.*[10]

In children with NF1, only two eye-tracking studies were conducted to investigate the saccadic system in this population and no consensus emerged from the results. While Lasker et al. (2003) found a disruption in the saccadic system, Krab et al. (2011) didn't find any difference between children with NF1 and TD.

More precisely, Lasker et al. (2003) used four different paradigms to study respectively elicit reflexive saccades, suppression of reflexive saccades, memory function and elicit volitional predictive saccades, in NF1 children aged from 6 to 11 years. For the visually guided saccade paradigm, they found an increase in latency for 20° and 30° target eccentricities and a decrease in accuracy for the 30° target eccentricity. NF1 children also made more errors than TD children in the antisaccade paradigm and exhibited more difficulties and greater latency in the predictive saccade paradigm for the 0.5 Hz frequency stimulus. Finally, for the memory-guided saccade paradigm, NF1 children demonstrated a longer time to make a saccade for each of the remembered targets and a higher number of premature responses than controls. There are two main limitations in this study: the small size of the sample (i.e., 10 NF1 children) and the participants' inclusion criteria (e.g., 5 out of 10 children met the diagnostic criteria for ADHD).

[10] The term saccadic system here does not include what is related to eye-movement control in reading since no studies to our knowledge were conducted in this field.

In comparison, Krab et al. (2011) evaluated the saccade adaptation of 53 children with NF1 with a classical backward saccade adaptation paradigm. During 3 sessions (i.e., baseline, adaptation, and extinction trials), children were asked to look at a single dot that moved from the left to the right visual field relative to the centre of the screen. The results revealed no significant differences between the NF1 group and the comparison group in terms of the baseline saccadic performance (i.e., number of correct primary saccades, baseline saccadic gains and variability) and the saccadic adaptation (i.e., size of the adapted gains and adapted saccade variability).

Given these divergent results, it is important to note that even if the two studies assessed the saccadic system in NF1 children, they (1) did not use the same paradigms and (2) did not deal with reading and the relevance of its relationship to eye-movement control.

***Visuo-attentional processes.***

To our knowledge, only one study, by our research team, has explored the relationship between visuo-attentional processes and reading in this population (Vernet  al., 2022b). This study used the DEM-test to evaluate the visuo-attentional scanning processes involved in reading. The results showed that visual-processing deficits were highly present in NF1 children. However, only comorbid poor readers presented an increased risk of visual-processing difficulties compared to their peers. This supports the implication of visuo-attentional deficits in the reading difficulties frequently observed in NF1 children.

Note also that several research studies used visual search tasks to investigate selective visual attention skills. The most common task used to this end was the *Sky search* subtest (*TEA-Ch*). With this task, Payne et al. (2011)[11] showed significantly lower selective visual attention outcomes for the NF1 group compared to unaffected siblings. Contrarily, Arnold et al. (2020), Hyman et al. (2005), Isenberg et al. (2013), Pobric et al. (2022) and Pride et al. (2018) didn't find significant differences between groups. This last result was corroborated in an experimental visual search task (Van Eylen et al., 2017) and the *Visual attention* subtest of the *NEPSY* with pre-schoolers (Lorenzo et al., 2013). Moreover, Vaucheret Paz et al. (2019) showed that 33.3% of NF1 children performed poorer in this last subtest (i.e., the *Visual attention* subtest of the *NEPSY*), but the sample performing this test was very limited (i.e., only 6 children). Regarding reading, Watt et al. (2008) reported no significant correlation between selective visual attention and reading subtests in NF1 children.

Visual selective attention was also studied through the *Stroop Color-Word Test (SCWT)* and the observed results were less controversial than with the visual search tasks. While Descheemaeker et al. (2005) showed that children with NF1 displayed average scores (i.e., mean score of -0.53 SD; see Krab et al., 2008 for a similar result), nearly half of those children (i.e., 53%, 9 children out of 17) obtained results below -1 SD in this task. Remigereau et al. (2018) found a significant difference between the NF1 children and the control group for the interference time, with a moderate effect size. Finally, the only study on academic achievement using this test showed that children with NF1 and general learning disabilities performed worse on the *SCWT* than those with NF1 and a specific learning disability (Krab et al., 2008). However, the study didn't provide the results according to each impaired academic domain. So, we cannot specifically define the link between difficulties in this task and reading deficits.

[11] Note that the 81 NF1 children of the study by Hyman et al. (2005) were also included in the 199 NF1 children of the study by Payne et al. (2011). Thus the divergent results of these two studies could be attributed to differences in the sample sizes: it is possible that the significant effect found in the Payne et al. (2011) study may be more representative of the NF1 population due to the large sample size of this study.

Finally, Michael et al. (2014) investigated elementary visuo-attentional components in children with NF1 in a visual reactivity task with and without distractors occurring at varying time intervals. This task allowed them to assess three attentional processes (i.e., alerting, distraction, and interruption). Although no significant differences were found for the alerting and distraction indices, this study revealed differences between the NF1 and control groups for the interruption index. The higher reaction times in NF1 children for this index could reflect a lower resistance to visual interference and thus a disturbance in the ability to focus visual attention on a specific target. This result is in line with the disruption of the alpha modulation found in EEG studies on tasks involving the overt and covert deployment of visual attention. Ribeiro et al. (2014) reported an increase in alpha oscillation amplitude in the parieto-occipital regions of the NF1 children compared to the control subjects both at rest and before poor performances in an overt attention task in which the target was centrally presented. Silva et al. (2016) also showed abnormal alpha modulations in children with NF1 when the attention was covertly allocated to a target outside of the central visual field. Specifically, in a target offset detection task, NF1 children achieved similar response accuracy as the control group but associated with a significantly higher desynchronization of alpha oscillations in NF1 children. This high desynchronization was interpreted by the authors as a compensatory mechanism necessary to obtain comparable performance to the control group and to compensate for a dysfunction of the visual attentional control processes.

## DISCUSSION

This systematic review aimed to examine the literature on visuo-perceptual (with and without a motor implication) and visuo-attentional processes in NF1 children, and to clarify their possible impact on the reading deficits frequently experienced by these children. The results revealed that NF1 children displayed very frequently impaired visual-processing skills (see Table 3 for a summary of the results). However, it is important to emphasize that contradictory results are frequently observed in studies. This limits the synthesis of results and thus the emergence of a clear conclusion regarding the effectiveness of these visual processes in children with NF1. It is also important to point out that many of the visual deficits seem to co-occur with reading deficits but are not functionally relevant to reading. The idea of this discussion is therefore to summarize the main results and to highlight the relationship that these visual processes may have with NF1 children's reading skills (functional role vs. co-occurrence).

**Significant weaknesses both in reading and visual processing skills.**

Many cognitive functions can be impaired in NF1 children (for reviews, see Crow et al., 2022; Lehtonen et al., 2013) and several of them can affect reading (e.g., linguistic skills; Arnold et al., 2018, 2020). In this framework, the issue of the visual processes' involvement in NF1 children's reading deficits has emerged through the study of visuo-spatial perception (Cutting et al., 2000; Cutting & Levine, 2010). In most of the selected studies, NF1 children exhibited significantly reduced performance than controls and a high frequency of deficits in this visual skill, whatever the test used to assess visuo-spatial abilities (JLO, Cutting & Levine, 2010; Thurstone test, Chaix et al., 2017; TVPS, Arnold et al., 2018; BORB, Hyman et al., 2005). The results also indicated that NF1 children with a reading deficit performed more poorly on visuo-spatial perception tasks than TD or RD children (Barquero et al., 2015; Cutting & Levine, 2010). A significant correlation between visuo-spatial processes and reading was also reported especially in NF1 patients with a reading deficit (Cutting & Levine, 2010). However, it is important to point out that as mentioned in the *Introduction* section, visuo-spatial perception processing was rarely directly studied to address the issue of reading acquisition in children with DD or TD, lessening the possibility of establishing a causal relationship between visual processing and reading deficits in NF1.

Among the motor functions supposed to be linked to reading, visual-motor integration and visual-constructive abilities were often found to be poorer in children with NF1 compared to TD children. Despite a wide variability between studies, the prevalence of perceptivo-motor difficulties in NF1 appears to be considerably higher than in the general population. However, few studies have attempted to link these visual-motor processes to reading and, as with visuo-spatial processes, there is some evidence suggesting that other cognitive processes may be involved in the tasks used, which could partly explain the failure to complete them and make these tasks unsuitable for demonstrating the role of visual-motor skills in reading failure.

To our knowledge, only two eye-tracking studies were already conducted on children with NF1 (Krab et al., 2011; Lasker et al., 2003). These two studies focused on the saccadic system and differed in their results. While Lasker et al. (2003) found evidence of saccadic computation differences between NF1 and TD children, Krab et al. (2011) did not observe any difference between groups. As noted in the *Results* section, they did not examine the same processes and the study of Lasker et al. (2003) contained many limitations regarding sample size and inclusion criteria. Furthermore, the two eye-tracking studies in NF1 children did not consider the children's reading level, nor study saccade computation during reading or reading-like tasks. But in any case, it is highly unlikely that eye movements per se are causally related to a reading disorder. Disruption to oculomotor behaviour often reflects the fact that many of the component processes involved in learning to read have not yet become fully automatized (Blythe et al., 2018; LaBerge & Samuels, 1974), probably related to a limited reading experience (see Cunningham & Stanovich, 1997; Goswami, 2015).

Some studies have suggested a delay in the maturation of low-level vision processes in children with NF1 (Lasker et al., 2003; Ribeiro et al., 2012). In that regard, alterations of the magnocellular-dorsal system were reported in children with NF1 (Ribeiro et al., 2012; Violante et al., 2012). The disruption in this visual pathway could result in difficulties (1) to reach a global precedence level of perceptive analysis, (2) to deploy efficient spatial attention and (3) to develop usual eye-movement control strategies, as would be done by expert readers (Stein, 2019, 2022). However, Ribeiro et al. (2012) failed to find a significant correlation between the magnocellular-dorsal system functioning and the visuo-spatial perception assessed by the JLO. Additionally, Van Eylen et al. (2017) did not find a significant difference between individuals with NF1 and TD children in a coherent motion task which engages the activation of the magnocellular pathway. In light of these conflicting results, two important points of discussion need to be stressed. First, the magnocellular hypothesis in dyslexia only addresses children with reading deficits. However, in all the studies with NF1 children, the children's reading proficiency was not considered. Secondly, as mentioned in the introduction, the causal hypothesis of a magnocellular deficit associated with visual-processing difficulties and reading deficits is highly controversial (Blythe et al., 2018; Hutzler et al., 2006). Although, as in dyslexia, some children with NF1 exhibit magnocellular dysfunction, its relationship to the reading deficits encountered by children with NF1 must be interpreted with caution.

To summarize, it is very common for children with NF1, including those with reading disabilities, to perform poorly on visuo-spatial and perceptivo-motor tasks, yet that does not mean that these visual deficits play a role in explaining the reading difficulties of NF1 children. Despite their high prevalence in NF1, these visuo-spatial or visual-motor deficits are more likely simple "*fellow travellers*"[12] of this disease, that co-occur with reading deficits.

**Role of visuo-attentional deficits in NF1 children's reading disabilities.**

The results on the deployment of attention in NF1 children highlighted lower resistance to visual interference and thus, difficulties to focus visual attention on a specific target (Michael et al., 2014). In other words, children with NF1 demonstrate difficulties in focusing attention on relevant information and inhibiting irrelevant information in covert attention, as already demonstrated in children with DD (Franceschini et al., 2012; White et al., 2019). This result is in line with the reported difference in alpha-band oscillations between children with NF1 and TD children in a covert attention task (G. Silva et al., 2016). This kind of attention is central since it is involved in parafoveal pre-attentive processing. It facilitates recognition of the currently fixed word and also influences saccadic computation towards the next fixation (for a review, see Schotter et al., 2012). Attentional focusing difficulties in NF1 children are consistent with those observed in DD children with a more diffuse distribution of attention (Facoetti et al., 2000, 2006). DD children displayed a lack of attentional inhibition for targets at the uncued location in the right VF (Facoetti et al., 2006). This abnormally important allocation of attentional resources (Facoetti et al., 2006; Facoetti & Turatto, 2000) is related to greater vulnerability to the crowding effect (Bellocchi, 2013; Martelli et al., 2009) and consequently, to a reduced parafoveal preview benefit in individuals with dyslexia compared to expert readers (S. Silva et al., 2016). In addition, as previously stated in the introduction, the difficulty in narrowing their focus of attention also hampers the exact planning of fine-tuned saccades (e.g., flattened, and diffuse landing position curve, unexpected/atypical saccades; e.g., Bellocchi et al., 2019; Facoetti, 2012; Facoetti et al., 2010; Lobier et al., 2012; Valdois et al., 2004; see Gavril et al., 2021 for a meta-analysis). Thus, in NF1 children, the attentional focus defect may lead to the same pattern of interference in parafoveal processing, with increased sensitivity to the visual crowding effect, reduced parafoveal pre-processing and occasional oculomotor control deficits, thus supporting the role of these visuo-attentional deficits in contributing to reading disabilities encountered in nearly half of the children with NF1.

[12] We thank Reviewer 1 for this suggestion.

A selective attention defect was also shown in visual information processing at the foveal level. In non-NF1 children, a relationship between reading achievement and visual target search (measuring the shift in the foveal allocation of attention) has been demonstrated (Franceschini et al., 2012; Guilbert & Guiraud- Vinatea, 2022). The more a child automates reading, the more precise he/she will be in his/her visual search with (1) less return to areas already explored, (2) more regular movements towards the nearest target and (3) a preferential exploration by following lines. In that sense, Vernet et al. (2022) used the DEM-test that provides an indirect measure of the efficiency of visual-attention processes in a simulated reading task (horizontal and vertical digit naming task). They showed that poor NF1 readers appeared to be at increased risk for visual-processing impairments compared to peers, with greater prevalence of failure on the horizontal digit naming subtest (related to a deficit in the visual-attention processes specifically involved in reading and more precisely to the left-to-right directionality of visual scanning). The poor left/right scanning abilities of these children could limit their reading speed and accuracy. In addition, this result could be explained by a reduction in the visual attention span, limiting the parallel processing of information as in DD children (Bosse et al., 2007; Prado et al., 2007). If this assumption is confirmed, it could account for the low reading speed and the irregular word reading difficulties frequently experienced by children with NF1 (e.g., Arnold et al., 2020; Biotteau et al., 2019).

This idea is consistent with global-to-local visual processing often affected in NF1 children (Barquero et al., 2015; Bulgheroni et al., 2019; Clements-Stephens et al., 2008; Cutting & Levine, 2010; Payne et al., 2017). When a TD child becomes an expert reader, he/she will progressively shift from a grapho-phonological decoding mode (local perceptual analysis) to lexical processing at the word level (Coltheart et al., 2001), with precedence of global level analysis (Krakowski et al., 2016; Poirel et al., 2008). Results from the present literature review suggested that like children with DD (Franceschini et al., 2017), NF1 children did not exhibit the global precedence bias over local processing. This perceptual analysis specificity could impact the orthographic processing strategy implemented by NF1 children. It may result in difficulties in the transition from grapho-phonological decoding of sub-lexical information to global word processing. Thus, the inability to prioritise a global processing strategy and to extend visual attention deployment in the foveal area appears to be closely related to the reading automation failure of some NF1 children.

**Perspectives.**

A major limitation in studies assessing visual processes in children with NF1 is that they did not consider their frequent comorbidity with reading deficits in the NF1 group (Descheemaeker et al., 2005; Hyman et al., 2006; Orraca-Castillo et al., 2014). Probably for this reason, the few studies that have attempted to link visual-motor process to reading failed to find a significant correlation between them in the sample of NF1 children (Mazzocco, 2001; Watt et al., 2008), while studies on non-NF1 children demonstrated a strong relationship between reading deficits and visual-motor difficulties, allowing for differentiation between TD and DD children (Bellocchi et al., 2017; Iversen et al., 2005; Kooistra et al., 2005) and to predict the reading level of children as early as in kindergarten (Bellocchi et al., 2017). Further studies considering this factor and dissociating NF1 children with and without a reading deficit would help to clarify this area of research.

In this context, the heated debate on the role of the visual system in reading needs more longitudinal studies starting in kindergarteners to increase the likelihood of being able to reliably identify explicit causal links. Studies with preschool-aged children make it possible to determine the prerequisites for reading acquisition and then, to identify at an early age those NF1 children at risk of presenting reading deficits. Much research on NF1 preschool-aged children has already been conducted to study cognitive functions and adaptative behaviour in kindergarten (Arnold et al., 2018; Beaussart et al., 2018; Brei et al., 2014; Klein-Tasman et al., 2013; Sangster et al., 2011; Soucy et al., 2012; Thompson et al., 2010), but no research has to our knowledge explored the role of early visual skills in the learning-to-read process. The design of longitudinal research protocols therefore appears to be the appropriate way to study the prerequisites specifically involved in reading difficulties in NF1 by following children from kindergarten to explicit reading acquisition. From a clinical viewpoint, specifying the role of visual skills in learning to read at an early age would make it possible to develop and implement remediation with clinical tools adapted to the reading profile of NF1 children and to put this in place before the reading delay becomes too great.

Studying the efficiency of visuo-perceptual and visuo-attentional processes in NF1 children according to the presence of reading deficit comorbidity requires the use of tools directly related to the visual processes of interest. The results from this literature review revealed the involvement of a range of confounding cognitive skills in many of the tools used. For instance, Van Eylen et al. (2017) identified that the differences observed in the *ROCF* copy did not remain significant after controlling for inhibitory control. In the same vein, the *JLO* test involves visual processes associated with cortical activation of the posterior regions (including parietal and occipital areas) and also executive abilities associated with additional activation within frontal regions (Clements-Stephens et al., 2008). Worse performance in the *SCWT* may be also interpreted as both visual selective attention and inhibitory control deficits. It is therefore difficult to isolate the attentional component of poor performance from the other skills involved in this task such as executive dysfunction (Van Eylen et al., 2017). In light of these observations, it seems essential to include unconfounded measures as covariates in the modelling analyses in order to limit their impact on the observed effects. Although it is hard to focus on a specific cognitive process in neuropsychological assessment from psychometric tests, there is a real challenge in assessing visual-processing skills specifically involved in reading. For instance, it would be interesting to evaluate left/right visual scanning, visual attention span, preferred level of perceptual analysis, or even eye-movement control and saccade targeting strategies using psychometric tests or experimental tools that have previously demonstrated their relevance to reading (e.g., Bellocchi et al., 2017, 2021; Bellocchi & Ducrot, 2021; Bosse et al., 2007; Ducrot et al., 2013; Facchin, 2021; Krakowski et al., 2016). Assessment tools for the visual-attentional processes specifically involved in reading difficulties could then be included in the recommendations for the assessment of attentional difficulties in NF1 children (Klein-Tasman et al., 2021; Pardej et al., 2022).

What it is clear is that eye-movement patterns are strictly linked to the visuo-attentional processes specific to reading behaviour and that these non-linguistic processes could serve as an additional source to explain impaired reading in NF1. In order to underline overlaps between NF1 and other learning disabilities, eye-movement recording could again be an excellent method. More studies directly comparing different learning disabilities (whilst taking account of the children's reading level) would provide invaluable information on their specificities or commonalities and on the occurrence of visual or visuo-attentional deficits in NF1. Given the great heterogeneity of cognitive profiles in NF1, it seems also important to specify the visual-processing characteristics according to the nature of the reading difficulties. Indeed, in DD, many studies report distinct visuo-perceptual and visuo-attentional difficulties regarding the affected reading procedure (i.e., graphophonological decoding of sub-lexical units or lexical processing; Bosse et al., 2007; Facoetti et al., 2006; Goldstein-Marcusohn et al., 2020). All these elements will allow a better understanding of whether the clinical profiles associated with reading difficulties in NF1 are specific to this disease or whether they are transposable to those observed in DD.

A better understanding of the visual processes involved in reading difficulties will allow us to provide clinical remediation tools adapted to the NF1 child's profiles. As already studied for phonological skills (Arnold et al., 2016; Barquero et al., 2015), specific remediations focused on visual processes need to be developed. In NF1 children, D'Archangel et al. (2022) provided preliminary results on a virtual maze showing that, despite lower initial performance, NF1 children were able to reach a level of performance equivalent to that of reading level-matched non-NF1 children, with additional practice. In the field of reading, several computerised tools based on visuo-attentional processes were proposed to improve reading skills in DD (e.g., Pasqualotto et al., 2022; see also Franceschini et al., 2015, for a review on action video games). For instance, visual attention span training can specifically improve the lexical reading process, by facilitating the parallel processing of the word’s letters (Valdois et al., 2014; Zoubrinetzky et al., 2019). An increase inter-letter and inter-word spacing and a reduced number of words per line also limit the effect of perceptual crowding, which is more pronounced in DD children (e.g., Perea et al., 2012; Schneps et al., 2013; Zorzi et al., 2012). Eye-movement control strategies and more precisely saccade targeting efficiency could also be improved with real-time feedback on the fixation position (Lehtimäki & Reilly, 2005) or the use of a brighter/coloured letter in a word to attract the eye towards the optimal viewing position and optimize the location of the initial fixation within the word (Ducrot et al., 2023; Vernet et al., 2023). All these clinical tools could be useful for the management of reading deficits in children with NF1.

To conclude, this literature review highlights weaknesses in the different components of the visual processes involved in reading in children with NF1. More specifically, attentional focusing dysfunction, poor left/right attentional scanning abilities, precedence of the local perceptual analysis strategy, visual-motor integration deficit and alteration of the visual magnocellular system have been demonstrated in children with NF1. While there are divergent results in the literature, these specificities could partly explain or reflect the highly frequent comorbid reading deficits in this disease. The results raise the importance of directly studying the visuo-perceptual, visuo-attentional processes and eye-movement control involved in the learning-to-read process in NF1. This discussion provides new directions for research, both for a better understanding of the cognitive mechanisms involved in this disease and also for the screening and care of learning disorders associated with NF1 to reduce school failures and their strong lifelong negative impact.

**Table 1.** Tests used in studies on NF1 children assessing visual processes involved in reading

| Visual processes assessed | Test | Subtest |
|---|---|---|
| Visuo-perceptual<br>(including global-to-local process) | DTVP/DTVP-2 | Figure-ground<br>Form constancy<br>Visual closure |
| Visuo-spatial perception | BORB | Gap matching<br>Line orientation |
| | DTVP/DTVP-2 | Position in space |
| | FKSB | Recognition-Discrimination |
| | HVOT | - |
| | JLO | - |
| | NEPSY | Arrows |
| | Thurstone test | |
| | TVPS / TVPS-R | Visual-spatial relations<br>Visual Discrimination |
| | WJ-R | Spatial relations |
| Visual-motor/visual-constructive | Beery VMI | Visual-Motor Integration<br>Motor coordination |
| | Bender Visual-Motor Gestalt Test | - |
| | DAS-II | Copying |
| | NEPSY/NEPSY-II | Design copying<br>Visual motor precision |
| | ROCF | - |
| Visuo-attentional | DEM-test | - |
| | NEPSY | Visual attention |
| | SCWT | - |
| | TEA-Ch | Sky search |

*Notes*. Only 2D tests were retained to be as close as possible to the conditions found in the reading. We acknowledge that one test does not isolate a specific process, it is therefore a schematic classification to highlight the main visual process that is evaluated. *DTVP* Developmental Test of Visual Perception, *BORB* Birmingham Object Recognition Battery, *FKSB* Florida kindergarten screening battery, *HVOT* Hooper Visual Organization Test, *JLO* Judgment of Line Orientation, *TVPS* Test of Visual Perceptual Skills, *WJ-R* Woodcock–Johnson–revised Test of Cognitive Abilities, *Beery VMI* Beery–Buktenica Developmental Test of Visual-Motor Integration, *DAS-II* Differential Ability Scales-II, *ROCF* Rey-Osterrieth Complex Figure Test, *SCWT* Stroop Color-Word Test, *TEA-Ch* Test of Everyday Attention for Children.

**Table 2.** Characteristics of the studies selected in this systematic review

| Authors (year) | Sample size (F/M) | Mean age (SD) | Task(s) |
|---|---|---|---|
| Arnold et al. (2018) | NF1: 42 (19/23)<br>TD: 32 (17/15) | NF1: 5.5 (0.5)<br>TD: 5.3 (0.5) | Visual-spatial relations (TVPS-R) |
| Arnold et al. (2020) | NF1: 60 (28/32)<br>TD: 36 (18/18) | NF1: 8.7 (1.8)<br>TD: 8.8 (1.6) | JLO<br>Sky search (TEA-Ch) |
| Barquero et al. (2015) | NF1+RD: 17 (6/11)<br>RD: 32 (15/17)<br>TD: 26 (13/13) | NF1+RD: 10.4 (1.5)<br>RD: 10.2 (1.9)<br>TD: 9.7 (1.5) | JLO<br>Visual closure (DTVP)<br>Position in space (DTVP) |
| Baudou et al. (2020) | NF1: 38 (23/15)<br>TD: 42 (20/22) | NF1: 9.1 (1.3)<br>TD: 9.5 (1.1) | JLO |
| Billingsley et al. (2002) | NF1: 24 (12/12)<br>TD: 24 (12/12) | NF1: 11.0 (N.A.)<br>TD: 11.8 (N.A.) | JLO<br>Recognition-discrimination (FKSB) |
| Bulgheroni et al. (2019) | NF1: 18 (7/11)<br>Siblings: 17 (8/9)<br>TD: 18 (7/11) | NF1: 10.2 (2.5)<br>Siblings: 12.2 (0.3)<br>TD: 10.2 (2.9) | ROCF |
| Casnar et al. (2014) | NF1: 38 (17/21)<br>TD: 23 (8/15) | NF1: 5.3 (0.7)<br>TD: 5.4 (0.8) | Copying (DAS-II)<br>Visual motor precision (NEPSY-II) |
| Chaix et al. (2018) | NF1: 75 (39/36)<br>TD: 75 (39/36) | NF1: 10.0 (1.3)<br>TD: 10.0 (1.2) | JLO<br>Thurstone test |
| Clements-Stephens et al. (2008) | NF1: 13 (6/7)<br>TD: 13 (6/7) | NF1: 9.8 (1.8)<br>TD: 9.8 (2.6) | JLO<br>HVOT<br>Position in space (DTVP-2)<br>Visual closure (DTVP-2) |
| Cutting and Levine (2010) | NF1+RD: 13 (4/9)<br>NF1noRD: 12 (7/5)<br>TD: 36 (14/22)<br>RD: 33 (10/23) | NF1+RD: 10.3 (2.2)<br>NF1noRD: 9.6 (2.1)<br>TD: 9.6 (2.3)<br>RD: 9.3 (1.2) | JLO<br>HVOT<br>Position in space (DTVP)<br>Visual closure (DTVP) |
| Cutting et al. (2000) | NF1: 20 (3/17)<br>ADHD: 13 (5/8)<br>TD: 16 (8/8) | NF1: 9.6 (2.4)<br>ADHD: 8.1 (2.1)<br>TD: 10.2 (2.5) | JLO |
| D'Archangel et al. (2022) | NF1: 17 (5/12)<br>TD: 14 (6/8) | NF1: 12.9 (3.3)<br>TD: 10.3 (2.7) | JLO |
| Descheemaeker et al. (2005) | NF1: 17 (5/12) | NF1: 9.2 (N.A.) | ROCF<br>Beery VMI<br>Bender Visual-Motor Gestalt Test<br>SCWT |
| Dilts et al. (1996) | NF1: 20 (12/8)<br>Siblings: 20 (12/8) | NF1: 10.8 (N.A.)<br>Siblings: 12.5 (N.A.) | JLO<br>TVPS<br>Beery VMI |
| Eldridge et al. (1989) | NF1: 13 (3/10)<br>Siblings: 13 (9/14) | NF1: 13.1 (6.4)<br>Siblings: 14.5 (5.0) | JLO<br>Beery VMI |
| Erdoğan-Bakar et al. (2009) | NF1: 27 (14/13)<br>Siblings: 20 (16/4)<br>ADHD: 40 (20/20)<br>TD: 40 (20/20) | NF1: 11.1 (F) / 10.5 (M) (N.A.)<br>Siblings: 13.2 (F) / 9.5 (M) (N.A.)<br>ADHD: 9.8 (F) / 10.2 (M) (N.A.)<br>TD: 10.1 (F) / 10.2 (M) (N.A.) | JLO<br>Bender Visual-Motor Gestalt Test |
| Gilboa et al. (2010) | NF1: 30 (21/9)<br>TD: 30 (21/9) | NF1: 12.2 (2.5)<br>TD: 12.3 (2.4) | Beery VMI |
| Gilboa et al. (2014) | NF1: 30 (21/9)<br>TD: 30 (21/9) | NF1: 12.2 (2.5)<br>TD: 12.3 (2.4) | JLO<br>Beery VMI<br>ROCF |
| Hofman et al. (1994) | NF1: 12 (2/10)<br>Siblings: 12 (6/6) | NF1: 10.4 (N.A.)<br>Siblings: 11.2 (N.A.) | JLO<br>ROCF |
| Hyman et al. (2005) | NF1: 81 (43/38)<br>TD: 49 (29/20) | NF1: 11.5 (N.A.)<br>TD: 12.0 (N.A.) | JLO<br>Spatial relations (WJ-R)<br>Gap matching (BORB)<br>Line orientation (BORB)<br>ROCF<br>Sky search (TEA-Ch) |
| Iannaccone et al. (2002) | NF1: 16 (7/9)<br>TD: 14 (N.A.) | NF1: 10.4 (2.8)<br>TD: 10.6 (3.6) | Visual evoked potential tasks with transient pattern-reversal and flash stimuli |
| Isenberg et al. (2013) | NF1: 55 (N.A.) | NF1: 9.7 (2.6) | JLO<br>Sky search (TEA-Ch) |
| Krab et al. (2008) | NF1: 86 (39/47) | NF1: 11.9 (2.5) | JLO<br>ROCF<br>Beery VMI<br>SCWT |
| Krab et al. (2011) | NF1: 70 (34/36)<br>TD: 19 (13/6) | NF1: 12.3 (2.5)<br>TD: 10.7 (2.1) | Beery VMI<br>Saccade adaptation task |
| Lalancette et al. (2022) | NF1: 28 (15/13)<br>TD: 28 (13/15) | NF1: 9.4 (2.4)<br>TD: 8.9 (2.4) | Steady‐ state visual evoked potentials with coloured icons flickering |
| Lasker et al. (2003) | NF1: 10 (3/7)<br>TD: 12 (5/7) | NF1: [6-11] [a]<br>TD: [6-12] [a] | Visually guided saccade paradigm<br>Antisaccade paradigm<br>Memory-guided saccade paradigm<br>Prediction paradigm |
| Lehtonen et al. (2015) | NF1: 49 (24/25)<br>Siblings: 19 (10/9) | NF1: 11.9 (3.2)<br>Siblings: 12.7 (2.7) | JLO |
| Lorenzo et al. (2013) | NF1: 43 (11/32)<br>TD: 43 (11/32) | NF1: 40.2 (0.7) [b]<br>TD: 40.2 (0.5) [b] | Beery VMI<br>Visual attention (NEPSY) |
| Mazzocco (2001) | NF1: 11 (7/4)<br>TD: 66 (46/20) | NF1: 6.3 (N.A.)<br>TD: 6.5 (N.A.) | Position in space (DTVP-2)<br>Form constancy (DTVP-2)<br>Figure-ground (DTVP-2)<br>Visual closure (DTVP-2)<br>Beery VMI |
| Mazzocco et al. (1995) | NF1: 19 (3/16)<br>Siblings: 19 (7/12) | NF1: 9.9 (2.4)<br>Siblings: 10.1 (2.5) | JLO<br>ROCF |
| Michael et al. (2014) | NF1: 20 (6/14)<br>TD: 20 (10/10) | NF1: 9.2 (1.9)<br>TD: 9.7 (1.8) | Experimental target detection task |
| Moore et al. (2000) | NF1: 52 (N.A.)<br>TD: 19 (N.A.) | NF1: 9.8 (3.6)<br>TD: 10.9 (2.7) | JLO<br>Beery VMI<br>Recognition-discrimination (FKSB) |
| Payne et al. (2011) | NF1: 199 (91/108)<br>Siblings: 55 (33/22) | NF1: 10.6 (2.3)<br>Siblings: 11.2 (2.0) | Sky search (TEA-Ch) |
| Payne et al. (2013) | NF1: 71 (30/41)<br>TD: 29 (15/14) | NF1: 10.50 (4) [c]<br>TD: 10.00 (5) [c] | JLO |
| Payne et al. (2017) | NF1: 30 (18/12)<br>TD: 24 (15/9) | NF1: 10.8 (2.4)<br>TD: 10.2 (1.8) | Modified Navon paradigm |
| Pobric et al. (2022) | NF1: 16 (7/9)<br>TD: 16 (7/9) | NF1: 13.0 (1.6)<br>TD: 13.3 (1.6) | Sky search (TEA-Ch) |
| Pride et al. (2018) | NF1: 19 (10/9)<br>TD: 18 (9/9) | NF1: 11.0 (2.8)<br>TD: 10.5 (2.5) | Sky search (TEA-Ch) |
| Remigereau et al. (2018) | NF1: 18 (6/12)<br>TD: 20 (6/14) | NF1: 10.4 (2.4)<br>TD: 10.4 (1.9) | SCWT |
| Ribeiro et al. (2012) | NF1: 19 (12/7)<br>TD: 33 (20/13) | NF1: 11.5 (2.5)<br>TD: 11.7 (2.3) | JLO<br>Experimental contrast sensitivity tasks activating parvocellular, koniocellular or magnocellular visual pathways |
| Ribeiro et al. (2014) | NF1: 17 (12/5)<br>TD: 19 (11/8) | NF1: 11.9 (2.3)<br>TD: 12.9 (2.6) | Visual detection task under overt attention in EEG |
| Roy et al. (2010) | NF1: 36 (18/18)<br>TD: 36 (18/18) | NF1: 9.6 (1.7)<br>TD: 9.6 (1.7) | Arrows (NEPSY) |
| Sangster et al. (2011) | NF1: 26 (9/17)<br>TD: 21 (10/11) | NF1: 5.2 (0.5)<br>TD: 4.7 (0.5) | Visual discrimination (TVPS-R)<br>Beery VMI |
| Silva et al. (2016) | NF1: 16 (12/4)<br>TD: 24 (16/8) | NF1: 13.9 (2.7)<br>TD: 13.5 (2.7) | Peripheral attentional targets experimental task under covert attention in EEG |
| Ullrich et al. (2010) | NF1: 10 (5/5)<br>Siblings: 6 (3/3) | NF1: 13.5 (2.3)<br>Siblings: 12.7 (1.7) | JLO |
| Van Eylen et al. (2017) | NF1: 39 (15/24)<br>TD: 52 (20/32) | NF1: 12.6 (3.1)<br>TD: 12.3 (2.6) | Coherent Motion task<br>Fragmented Object Outlines task<br>Visual search task<br>ROCF |
| Vaucheret Paz et al. (2019) | NF1: 24 (10/14) | NF1: 9.9 (N.A.) | ROCF<br>Design copying (NEPSY)<br>Visual attention (NEPSY) |
| Vernet et al. (2022b) | NF1: 42 (20/22)<br>TD: 42 (26/16) | NF1: 9.9 (1.5)<br>TD: 10.0 (1.1) | DEM-test |
| Violante et al. (2012) | NF1: 15 (9/6)<br>TD: 24 (13/11) | NF1: 11.7 (2.9)<br>TD: 12.0 (2.3) | Experimental fMRI fixation task activating parvocellular or magnocellular visual pathways |
| Watt et al. (2008) | NF1: 30 (13/17) | NF1: 9.3 (1.3) | JLO<br>Beery VMI<br>Sky search (TEA-Ch) |

*Notes*. *N.A.* not available, *ADHD* Children with attention deficit hyperactivity disorder, *RD* Children with idiopathic reading disabilities, *NF1+RD* Children with NF1 and reading deficits, *NF1noRD* Children with NF1 without reading deficits, *TD* Typically developing children.

[a] Age range, [b] Age in months, [c] Median age (IQR)

**Table 3.** Schematic summary of the NF1 children's performance compared to the comparison groups for the different visual processes of interest

| Study | Visual process/system of interest | | | | | | Relation to reading studied |
|---|---|---|---|---|---|---|---|
| | Visuo-spatial | Global-local | Visual pathways | Visuo-attentional | Visual-motor | Saccadic | |
| Arnold et al. (2018) | – | | | | | | Yes |
| Arnold et al. (2020) | – | | | ø | | | Yes |
| Barquero et al. (2015) | – | – | | | | | Yes |
| Baudou et al. (2020) | – | | | | | | No |
| Billingsley et al. (2002) | ø | | | | | | No |
| Bulgheroni et al. (2019) | | – | | | – | | No |
| Casnar et al. (2014) | | | | | – | | No |
| Chaix et al. (2018) | – | | | | | | Yes |
| Clements-Stephens et al. (2008) | – / ø | – | | | | | Yes |
| Cutting and Levine (2010) | – / ø | – | | | | | Yes |
| Cutting et al. (2000) | ø | | | | | | Yes |
| D'Archangel et al. (2022) | ø | | | | | | Yes |
| Descheemaeker et al. (2005) | | | | – | – | | No |
| Dilts et al. (1996) | – | | | | – | | No |
| Eldridge et al. (1989) | – | | | | ø | | No |
| Erdoğan-Bakar et al. (2009) | – | | | | ø | | No |
| Gilboa et al. (2010) | | | | | – | | No |
| Gilboa et al. (2014) | – | | | | – | | No |
| Hofman et al. (1994) | – | | | | – | | No |
| Hyman et al. (2005) | – | | | ø | – | | No |
| Iannaccone et al. (2002) | | | – | | | | No |
| Isenberg et al. (2013) | – | | | ø | | | No |
| Krab et al. (2008) | – | | | ø | – | | Yes |
| Krab et al. (2011) | | | | | – | ø | No |
| Lalancette et al. (2022) | | | – | | | | No |
| Lasker et al. (2003) | | | | | | – | No |
| Lehtonen et al. (2015) | – | | | | | | No |
| Lorenzo et al. (2013) | | | | ø | – | | No |
| Mazzocco (2001) | ø | ø | | | ø | | No |
| Mazzocco et al. (1995) | – | | | | – | | Yes |
| Michael et al. (2014) | | | | – | | | No |
| Moore et al. (2000) | – / ø | | | | ø | | No |
| Payne et al. (2011) | | | | – | | | No |
| Payne et al. (2013) | – | | | | | | No |
| Payne et al. (2017) | | – | | | | | No |
| Pobric et al. (2022) | | | | ø | | | No |
| Pride et al. (2018) | | | | ø | | | No |
| Remigereau et al. (2018) | | | | – | | | No |
| Ribeiro et al. (2012) | – | | – | | | | No |
| Ribeiro et al. (2014) | | | – | – | | | No |
| Roy et al. (2010) | ø | | | | | | No |
| Sangster et al. (2011) | ø | | | | ø | | No |
| Silva et al. (2016) | | | | – | | | No |
| Ullrich et al. (2010) | – | | | | | | No |
| Van Eylen et al. (2017) | | ø | ø | ø | ø | | No |
| Vaucheret Paz et al. (2019) | | | | – | – | | No |
| Vernet et al. (2022b) | | | | – | | | Yes |
| Violante et al. (2012) | | | – | | | | No |
| Watt et al. (2008) | – | | | – | – | | Yes |

*Notes.* The results are schematically represented by the symbol " –" when they are in favour of NF1 children’s difficulties in the visual skill concerned. A result in favour of difficulties in the area of interest may correspond for instance to a significant difference with the comparison group or to a large prevalence of children with difficulties in this domain. In contrast, the symbol " ø" indicates results that are not different from those of the comparison group. Sometimes studies are displayed with both symbols. This means that two different tests assessing the same visual process give divergent results.

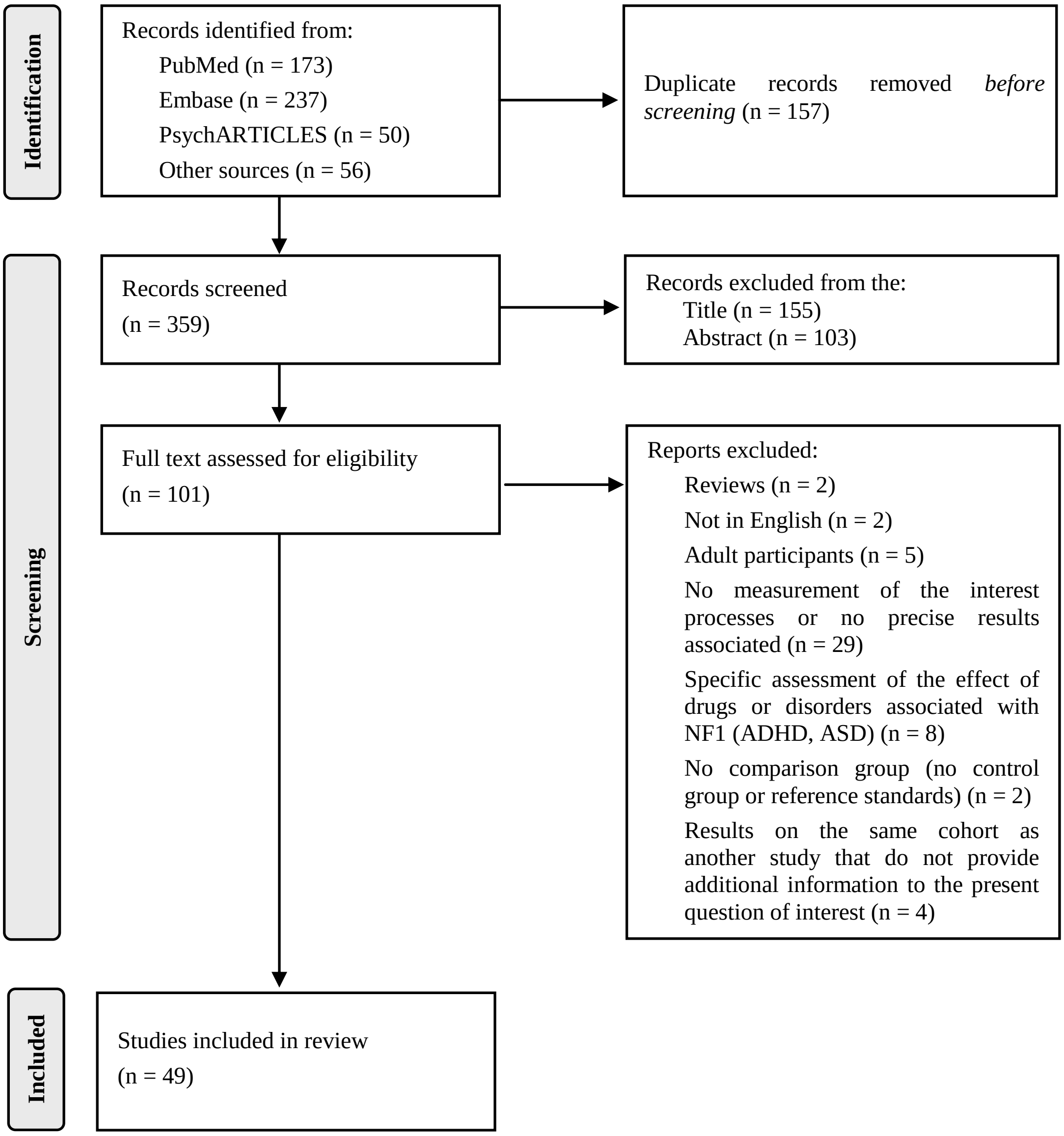

**Figure 1.** Flow diagram of the study selection process for the present systematic review